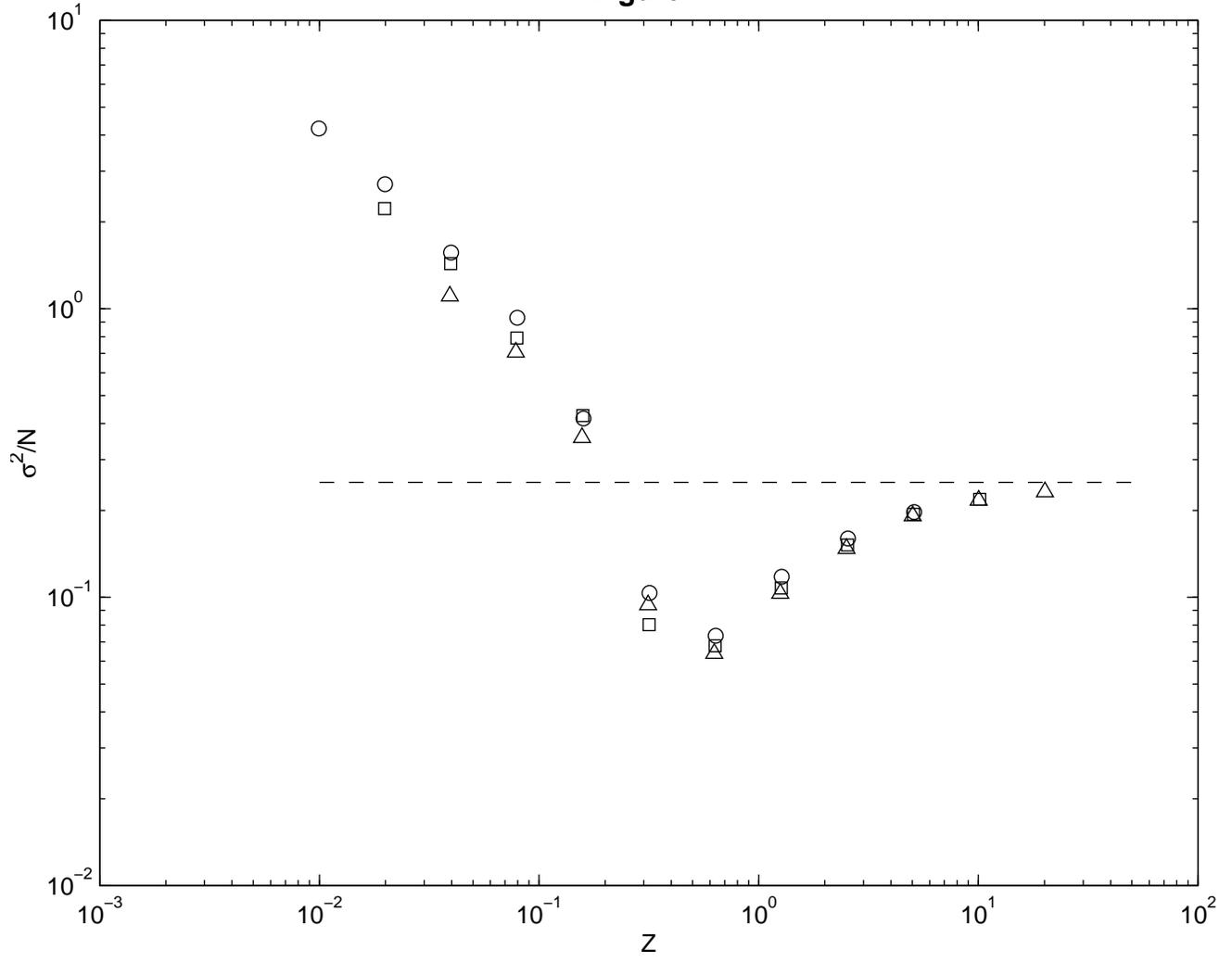
**Figure 1**

# The Minority Game with Variable Payoffs

Yi Li*, Adrian VanDeemen[+] and Robert Savit*

*Physics Department and Center for the Study of Complex Systems
University of Michigan, Ann Arbor, MI 48109
[+]University of Nijmegen, Nijmegen, Netherlands

Abstract

In the standard minority game, each agent in the minority group receives the same payoff regardless of the size of the minority group. Of great interest for real social and biological systems are cases in which the payoffs to members of the minority group depend on the size of the minority group. This latter includes the fixed sum game. We find, remarkably, that the phase structure and general scaling behavior of the standard minority game persists when the payoff function depends on the size of the minority group. There is still a phase transition at the same value of z, the ratio of the dimension of the strategy space to the number of agents playing the game. We explain the persistence of the phase structure and argue that it is due to the absence of *temporal* cooperation in the dynamics of the minority game. We also discuss the behavior of average agent wealth and the wealth distribution in these variable payoff games.

# I. Introduction

The minority game is a game played with heterogeneous agents which seeks to model a situation in which agents adaptively compete for a scarce resource.[1] In the simplest version of this model, each agent, at each time step of the game joins one of two groups, labeled 0 and 1. Each agent in the minority group is rewarded with a point, while each agent in the majority group gets nothing. An agent makes his decision about which group to join by using one of several strategies in his strategy set. The strategies use as input a set of common information. This model has a most remarkable structure:[2,3] Most notably, the system has a phase transition as a function of the ratio of the amount of information used the by the agents' strategies to the number of agents playing the game. At the phase transition the system also evidences an emergent coordination among the agents' choices which leads to an optimum utilization of the scarce resource, in a well-defined sense. This transition is remarkably robust, and persists even when the nature of the information used by the agents changes,[4] when the information becomes exogenous rather than endogenous,[5] and when the agents' strategies are allowed to evolve over time.[6]

In most studies of the minority game, the payoff to the agents is very simple, namely, each agent that is in the minority group is awarded one point, and each agent that is in the majority group is awarded nothing. In this paper we will consider an important modification to the game. We will study games with various payoffs to the agents (and attendant modifications of the rankings of the agents' strategies). In real social and biological systems, there may be various kinds of rewards for being in the minority, or for being innovative. It is easy to imagine situations in which the members of the minority group share a fixed amount of resource (a fixed-sum game): For example, a number of baboons exploring a seldom-traveled part of the forest, find a tree with some fruit. The amount of fruit discovered is fixed, so that the smaller the group, the more fruit there is for each member. Note that in this story, there would be no payoff at all if the baboons were not a minority, since the tree is in a "seldom-traveled" part of the forest. It is furthermore not difficult to generate scenarios in which total resources available to a minority group increase or decrease in various ways as the size of the minority group decreases, resulting in payoffs to members of a minority group that depend, in various ways on the size of the minority group. For example, a prospector may find a particularly rich mine (heretofore unknown), but his ability to mine the ore will be very limited unless he enlists the aid of partners. Working together they may be able to mine more per miner than each one could separately. This example also indicates that the issue of variable payoffs in the minority game is associated with the more general question of cooperation versus competition at different levels of organization. In the case of the miner, cooperation with his mining partners accounts for the increasing payoff with increasing size of the minority group, while competition with other miners in the discovery of the mine leads to the possibility of any payoff in the first place.

Aside from their interest for specific social and biological systems, minority games with different payoff functions are interesting from the point of view of the fundamental dynamics of emergent cooperation. In this context, games in which an individual's payoff increases as the size of the minority group decreases are particularly interesting. Studies of the minority game, heretofore, have shown that the dynamics of the game leads to situations in which the size of the minority group is maximized, so that, in some sense, social utility is maximized. This was most clearly demonstrated when evolution was added to the game. If one imposes payoffs that explicitly favor the formation of small minority groups, it is not *a priori* clear how the dynamics will proceed, or whether the phase transition will persist.

In the next section we will briefly review the minority game and describe the modifications to it that we shall study. In Section III we shall present our results, including results on overall resource utilization, as well as results on agent wealth and the wealth distribution among agents. We will find, surprisingly, that even with payoff functions that favor the formation of small minority groups, the fundamental phase structure of the game as a function of the amount of information available to the agents is unchanged: There is still a phase transition at the same critical parameter value as in the standard minority game, and at that transition the size of minority groups are maximized. Other measures of utility such as mean agent wealth and the wealth distribution among agents vary in an understandable way. The paper ends with a summary and discussion in Section IV.

## II. Description of the Game and its Modifications
### A. General Features of the Standard Minority Game

The minority game is a simple competitive game. The rules of the game are as follows: At each time step of the game, each of N agents playing the game joins one of two groups, labeled 0 or 1. Each agent that is in the minority group at that time step gains some reward, while each agent belonging to the majority group gets nothing. In the commonly studied version of the game, the reward for each agent in the minority group is fixed at one point, independent of the size of the minority group. An agent chooses which group to join at a given time step based on the prediction of a strategy. A strategy is a look-up table which returns a prediction of which group will be the minority group during the current time step of the game, given the value of a piece of data. In general, different agents use different strategies, but the data used by the strategies is commonly available to all the agents. That is, at each point in the game, each agent receives the same information (signal), and his strategies respond to that signal with a prediction. In the original game, the signals given to the agents are the list of m 0's and 1's that denote which were the minority groups for the past m time steps. Thus, a strategy of memory m is a look-up table of $2^m$ rows and 2 columns. The left column contains all $2^m$ combinations of m 0's and 1's, and each entry in the right column is a 0 or a 1. To use a strategy of memory m, an agent observes which were the minority groups during the last m time steps of the game and finds that entry in the left column of the strategy. The corresponding entry in the right column contains that strategy's prediction of which group (0 or 1) will be the minority group during the current time step. In this version of the game, the strategies use (the publicly available) information from the historical record of which group was the minority group as a function of time. Other versions of the game have been studied, including the interesting case in which the signals provided to the agents are exogenous IID integers between 0 and $2^m-1$.[5] The gross phase structure of the system is the same in these cases. In all cases, the controlling variable is $z \equiv D/N$, where D is the dimension of the strategy space from which the agents draw their strategies. (I.e., D is the number of different values that the publicly available signal can take on.) In the case in which the agents are given the m-history of the minority groups, $D=2^m$. In Fig. 1 we present a graph which shows how the system behaves as a function of z. The vertical axis is $\sigma^2/N$, where $\sigma$ is the standard deviation of the number of agents belonging to (say) group 1. A small value of $\sigma$ means that the minority groups are typically large, so that the scarce resource (membership in the minority group) is used near its limiting value. It is clear that as a function of z, the best resource utilization occurs at $z=z_c \approx 0.35$, which is also the point at which the system evidences a phase change. Note also the remarkable scaling properties of these results.[2,3]

**B. The Minority Game with Variable Payoffs**

The robustness of the general phase structure of the minority game is remarkable. But the payoff structure of the standard minority game is neutral with respect to the size of the minority group. It is not clear whether the same structure will obtain if the payoff functions are changed. In particular, it is not clear what the nature of the phase structure will be if the payoff structure to individuals does not favor the formation of large minority groups. In a fixed sum game, or in a game in which individual payoffs increase more rapidly as the size of the minority group decreases, we might expect qualitative changes in the phase structure of the system.

To address this question, we will study games in which we alter both the payoff to the agents, as well as the way in which the strategies are ranked. For simplicity, we will consider games in which the information used by the strategies is, as in the standard game, the list of which were the minority groups for the past m time steps. In the games we shall discuss, all strategies used by all the agents have the same value of m. At the beginning of the game each agent is randomly assigned s (generally greater than one) of the $2^{2^m}$ possible strategies of memory m, with replacement. For his current play the agent chooses from among his strategies the one that currently has the highest rank. In the original version of the model, a strategy is awarded one point for every time step in the past in which that strategy would have predicted the correct minority group.[7] In the cases we shall consider, the payoffs will generally depend on the size of the minority group. So, a strategy that would have successfully predicted the minority group at a given time step will be rewarded by an amount equal to the payoff given to each of the agents in the minority group at that time step.[8] Following each round of decisions, the cumulative performance of each of the agent's strategies is updated by comparing each strategy's latest prediction with the current minority group. Because each agent has more than one strategy, the game is adaptive in that agents can choose to play different strategies at different moments of the game in response to changes in their environment; that is, in response to new entries in the time series of minority groups as the game proceeds. Although this system is adaptive, the versions we analyze here are not, strictly speaking, evolutionary. The strategies do not evolve during the game, and the agents play with the same s strategies they were assigned at the beginning of the game.

To cover a range of possibilities, we shall consider both power law and exponential payoff functions. Let n be the population of the minority group, and let r=n/N. Then, we consider payoff functions of the form $A(r) \sim r^{-\alpha}$, and $A(r) \sim e^{-\gamma r}$. Here A(r) is the award made to each member of the minority group[9] when the minority group has a population, n=rN. The special case $\alpha$=1 is the fixed sum game, while the limits $\alpha \rightarrow 0$ or $\gamma \rightarrow 0$ recover the standard minority game. In addition to examining the behavior of overall resource utilization, $\sigma$, we will also study the wealth distribution among agents by looking at mean agent wealth and measures of the width of the wealth distribution. We choose to let A depend on r rather than on n since the results have a nicer scaling behavior, as we shall explain below. It is obvious that the results of games in which A depends explicitly on n can be obtained from our results by a simple rescaling of overall normalization or of γ.

## III. Results

Each of the simulations reported in this paper were run for 10,000 time steps. Values of N used were 51, 101, and 201 agents, and eight realizations of each experiment were preformed. To begin, we present, in Fig. 2a-e, $\sigma^2/N$ as a function of $z=2^m/N$ for several different games with different pay-off functions. Fig. 2a is the result for the game with exponential payoff and γ=4.0, thus strongly favoring the formation of small minority groups. Fig. 2b is the result for the game with power law payoff and α=2, also favoring small minority groups, and in particular, very strongly favoring the formation of very small minority groups. Fig. 2c shows the result for the fixed sum game, $A(r)=r^{-1}$. Figs. 2d and 2e show the result for games which weakly and strongly favor formation of large minority groups, namely, $A(r)=r^2$ and $A(r)=e^{0.1r}$, respectively. In all these graphs we see the same general behavior of $\sigma^2/N$ as a function of z that we saw in the standard minority game, Fig. 1. In particular, as in the standard game, all results appear to fall on the same universal scaling curve, and there appears to be a phase change in the system at the same value of z as in the standard game. At the phase transition we have the best overall resource utilization in the sense that the typical minority group is largest. Note also the remarkable fact that the actual values of $\sigma^2/N$ do not seem to depend on the form of the payoff in these graphs. The only exception to this is that $\sigma^2/N$ in the standard game is somewhat lower for small enough z than in the games with other payoff functions. We shall explain this effect below. This same picture obtains for games with more than 2 strategies per agent. Here also, the general behavior of $\sigma^2/N$ as a function of z does not materially depend on A(r).

To check further that the nature of system and its phase change is as in the standard minority game, we have also computed the conditional probabilities $P(1|u_m)$, which are the probabilities to have the minority group be group 1 following some specific string, $u_m$, of m minority groups, in the game played with memory m. As in the standard game, we find that for $z<z_c$ the histogram of $P(1|u_m)$ is flat at 0.5 for all $u_m$, while for $z \geq z_c$ it is not flat. This verifies that the phase transition in the games with different A(r) is of the same type as in the standard minority game, and may be thought of as a transition from an informationally strategy-efficient phase to a strategy-inefficient phase.[2,3]

In Figs. 3a-e, we plot $\overline{w}$, the mean wealth of the agents, as a function of z for the five cases studied in Figs. 2a-e. For comparison, we plot in Figs. 3f $\overline{w}$ for the standard minority game. In each of these figures we present results for several different values of N. We see some N dependence in most of the figures (with the exception of Fig. 3c, which is the fixed sum game, and for which the curve is flat and N independent by construction). It does seem that, as N increases, $\overline{w}$ as a function of z may be approaching a fixed curve. That is particularly apparent for $z>z_c$. It also appears that $\overline{w}$ is a minimum in Figs. 3a and 3b, and a maximum in Figs. 3d-f, even for large N. This is what we expect, since the minority groups are largest at $z=z_c$. However, an asymptotic curve which is flat, at least for $z \geq z_c$ cannot be ruled out.[10] The general shapes of the curves in Fig. 3 are easy to explain. The average agent wealth should be on the order of the payoff times the typical number of agents in the minority group. The typical number of agents in the minority group is closely related to σ, since the game is symmetric with respect to interchange of group 0 and group 1.[11] As a measure of the typical size of the minority group, therefore, we take $\overline{n} \approx \frac{N}{2} - \sigma$. Defining $\overline{r} = \frac{\overline{n}}{N}$ we plot as solid lines in Figs. 3 the quantity $\overline{r}A(\overline{r})$ as a function of z. We see that the curves match the simulation results nicely.

Another important question to address is the wealth distribution among the agents. There are a number of different measures that can be used to characterize the extent to which wealth is widely distributed among the agents, including the Gini coefficient, a measure common in the social sciences. Here we choose to look at η, the standard deviation of the wealth distribution among the agents. A small standard deviation means that the wealth is relatively broadly distributed, while a large standard deviation means that the wealth is concentrated in fewer agents. In Figs. 4a-e we plot the quantity $\tilde{\eta} \equiv \frac{\eta}{\overline{w}}\sqrt{N}$ for the games of Figs. 2 and 3. We also plot, in Fig. 4f, $\tilde{\eta}$ for the standard game. First, note that this quantity has a dip in the region near $z_c$ followed by a peak. The precise position of the dip and subsequent peak is somewhat dependent on A(r). Second, although these curves appear to scale fairly well, we do not believe that there is a simple universal scaling for η above and below the transition. In the first place, it is easy to show that, in the random choice game, (i.e. the game in which agents choose to join group 0 or group 1 independently and with equal probability), $\eta/\overline{w}$ is independent of N for large N. Indeed, in all the games studied here, this quantity is independent of N for large z. But $\eta/\overline{w}$ is not independent of N for other values of z. To see this refer to Figs. 5, in which we plot $\eta/\overline{w}$ as a function of z for various N. We have studied various possible scalings of $\eta/\overline{w}$ multiplied by different functions of N, and we have seen that a universal, z independent scaling form does not exist. Rather, it is possible, for a given payoff function A(r), to find a function $\phi_A$(N;z), such that $\frac{\eta}{\overline{w}}\phi_A(N;z)$ does become independent of N for large N.

It is also interested to note that for all those functions, A(r), that increase with decreasing r (thus favoring the formation of small minority groups), the curve of $\tilde{\eta}$ has a local minimum at z=$z_c$, while for those functions, A(r), which are not decreasing with increasing r (including the standard minority game), the curve of $\tilde{\eta}$ has a local minimum at a value z>$z_c$.

Changing the payoff function does not seem to materially affect the overall utilization of resources, in the sense that the curve of $\sigma^2$/N as a function of z is not substantially different than in the standard minority game. And while different payoff functions do alter details of agent wealth such as the mean of wealth and the nature of its distribution among agents, these changes are easily understandable, given the fact that $\sigma^2$/N as a function of z is not altered, as we shall explain in the next section. To test further the notion that changing the payoff to the agents does not change the fundamental phase structure of the system, we have constructed an artificial payoff function that strongly favors the formation of minority groups of a specific size. In particular, we consider a game played with N=101 agents and the payoff function A(r) = 1 for r ≠ $r_o$, and 10 for r=$r_o$. In Fig. 6 we plot $\sigma^2$/N as a function of z for several different choices of $r_o$ corresponding to favored minority groups of size $n_o$=49, 45 and 43. Again, this curve is substantially the same as in the standard minority game, and in particular, shows a minimum and a phase change at z=$z_c$.[12]

## IV. Discussion of the Results

The most significant result of this paper is the observation that the behavior of $\sigma^2$/N as a function of z is not materially altered when the payoff function is changed. $\sigma^2$/N still shows the same remarkable scaling behavior and phase structure that it had in the original, standard minority game.[2] One might have supposed that the behavior of $\sigma^2$/N would have been strongly affected by varying A(r), and in particular, by allowing A(r) to be large for small r. Then, it should have been possible for agents to arrange their choices so that an average agent would have been in the majority group well over 50% of the time, but when in the minority group that minority group would be small. In principle, this can lead to greater agent wealth (and high ranking for the appropriate strategy) if the payoff for small minority groups is sufficiently high. It is somewhat surprising that this does not happen, and that the curve of $\sigma^2/N$ as a function of z is largely unchanged (with the exception of very low z, an effect which we shall explain in a moment).

One way to understand this, is to recognize that an agent's choice about which group to join is dictated by that strategy that has the highest rank. For values of z not too far from $z_c$, (and following an initial learning period) most agents have one strategy which has a very high rank, and is used most of the time. Recall that the rankings of the strategies are cumulative over the course of the game, so such rankings represent average strategy performance.[13] Thus, the solution which the system discovers, and which results in large minority groups regardless of A(r), is based on the average performance of the agents' strategies. However, in order for the system to take advantage of A(r)'s that strongly favor small minority groups, requires *temporal* coordination. That is, agents must cooperate in the sense that agents must sacrifice themselves so that they place themselves in the majority group most of the time, and in the minority group only seldom. Given the way in which strategies are evaluated, it is fairly clear that such temporally structured solutions will not, in general, be found. One might suppose that the low z regime of the minority game contradicts this argument. Here the minority groups are, on average, relatively small. Furthermore, if A(r) favors small minority groups, it is possible for average agent wealth to exceed the value it has for larger z, as we see, for example, in Fig. 3a. However, the dynamics in this phase can hardly be said to be cooperative. Indeed, those agents who are wealthy are precisely those agents whose (two) strategies are most similar.[14] An agent who has two strategies whose responses to a given string, say $u_m$, are the same will be in the (small) minority group that follows $u_m$ about half the time. An agent whose two strategies respond differently to that string, though, will generally not be in the minority group following $u_m$. The kind of temporal cooperation necessary to take advantage of an A(r) that favors small minority groups is much closer to the successful strategies used in the iterated prisoner's dilemma, and related games.[15] There, players explicitly use the information that the game will continue in order to develop strategies that are sub-optimal locally in time, but perform much better over time. The question of modifying the minority game to allow agents to develop temporally coordinated strategies in order to generate mutually beneficial small minority groups will be discussed elsewhere.

In the context of a statistical mechanical description of the minority game, one might suppose that A(r) is something like an irrelevant operator. In ordinary statistical mechanical systems, irrelevant operators do not affect the critical exponents at second order phase transitions. However, our numerical results suggest that much more of the structure of the minority game is unaffected by changes in A(r), in particular, the shape of the curve of $\sigma^2$/N as a function of z and the value of $z_c$. In the context of a field theoretic approach to statistical mechanics, A(r), therefore, appears to play the role of a total derivative operator, the addition of which does not affect the dynamics of the system. If this is the case, then it should be possible to transform the minority game played with a non-trivial A(r) into the standard minority game by a canonical transformation.[16] On the other hand, it is unlikely that the phase transition observed at z=$z_c$ is a simple second order phase transition. It is more likely that the minority game is analogous to a spin-glass, and that the phase transition is analogous to that seen in spin-glass-like systems.[17] One must be cognizant of this essential complication. Of course, if the dynamics of the minority game were generalized sufficiently to allow for the temporal coordination described above, then A(r) would presumably cease to be a total derivative.

Although in general, the curve of $\sigma^2$/N as a function of z is independent of A(r), there is a deviation at small z for the standard game: In the standard game in which each strategy is awarded one point whenever it would have predicted the correct minority group, $\sigma^2$/N is less than for other choices of A(r) for low enough z. This anomalous result is easy to understand. As described in detail elsewhere,[3] for small enough z, the dynamics of the minority game can be well approximated by considering the response of the system to each m-string of minority groups as independent. As a consequence, for low enough z, the game is dominated by a period-two dynamics in which even occurrences of a given m-string of minority groups, $u_m$, give rise to the opposite minority group than occurred following the preceding odd occurrence of the same string, $u_m$. In the standard minority game, after each pair (odd-even) of occurrences of a given string, $u_m$, two strategies belonging to a given agent, but which have different predictions in response to $u_m$ will be tied in their ranking. Thus, at the next odd occurrence of $u_m$ such an agent will choose randomly between these two strategies, and the resulting size of the minority group will be close to 50% of the agents, within random fluctuations. On the other hand, even occurrences of $u_m$ will typically give rise to very small minority groups.[3] For other functions, A(r), rewards to the strategies depend on the size of the minority group. They are real numbers, not integers, and so ties in ranking between strategies will almost never occur. Thus, the minority group following a given string, $u_m$, will simply alternate between 0 and 1,[18] and the minority groups will all be small (on the order of the size of the minority groups following an odd occurrence of $u_m$ in the standard game). Consequently, the average size of the minority group is anomalously large in the standard game, and so $\sigma^2/N$ is anomalously small in the standard game, for small enough z.

Although $\sigma^2/N$ is largely independent of A(r) and scales with N, other derivative properties of the system such as average agent wealth and the wealth distribution do depend on A(r), and do not scale with N in a very simple way. In the last section, we showed that average agent wealth as a function of z was easily understandable in terms of $\bar{r}$. For those games in which A(r) favors small minority groups strongly enough[19], average agent wealth is maximal at $z_c$, while for those game in which A(r) increases with large minority groups, average agent wealth is smallest at $z_c$. It is also noteworthy that the wealth distribution has a local minimum near $z_c$. Although the wealth distribution is smaller for very large z, agent wealth is also considerably smaller there, so most agents may be said to be "equally poor", which is not surprising for values of z for which average dynamics is similar to the random choice game. But for those payoff functions which do not decrease with decreasing size of the minority group, there is some sense in which system-wide utility is maximized near $z_c$: Agents are on average wealthy, and the wealth is distributed relatively widely in the population. On the other hand, it is interesting that for somewhat larger values of z the situation is rather different, and there is peak in the wealth distribution curve so that agents may still be somewhat wealthy, on average, but the wealth is distributed less uniformly. We believe that this curve actually reflects a cross-over between the two different dynamics that obtain above and below $z_c$, although the detailed nature of the cross-over is not entirely clear to us.

In the last section, we also noted that there was significant N dependence in $\bar{w}$ and $\eta$ at fixed z. Consider, for example, the low z results of mean wealth in Fig. 3. Here we see that if A(r) favors small minority groups, mean wealth is larger for smaller values of N at the same z, whereas if A(r) favors large minority groups, mean wealth is smaller for smaller N at the same (low) value of z. These results can be traced to the fact that there is a finite size effect at low z, which results in average minority groups that are somewhat smaller for finite N, and, for fixed z, approach their asymptotic limit from below. This effect, to first order was explicitly calculated for the standard game at low z in Ref. 3. Although the detailed dynamics for arbitrary A(r) are somewhat different, the origin of the effect is still the same. Similar reasoning can be used to explain the other finite size effects in the mean wealth and the wealth distribution.

## V. Conclusion and Summary

In this paper we have shown that the main results of the minority game, in particular, $\sigma^2/N$ as a function of z, is materially independent of the payoff function to the agents. The system continues to have a phase transition at the same value of z, and the two phases have the same characteristics. We also discussed the behavior of average agent wealth and the wealth distribution, and showed that those functions were understandable given the payoff function and the fact that $\sigma^2/N$ as a function of z is materially the same as in the standard game. Finally, we discussed what we believe to be the fundamental dynamical reason for the persistence of this result, namely, the absence of nontrivial temporal cooperation among the agents.

Our results provide another important piece of information that points to the surprising robustness and universality of the phase structure of minority dynamics, at least for those systems in which there is an absence of temporal cooperation. And because different social and biological systems manifest different payoffs, our results strongly suggest the relevance of these dynamics to a wide range of complex adaptive systems. Indeed, in very simplified terms, different payoff functions, A(r), can arise from different detailed cooperative dynamics within groups. Thus, these games can be considered to be very simple models that capture some of the consequences of cooperation within groups. Understanding, in more detail, how minority dynamics manifests itself in specific social and biological systems is likely to lead to a deeper understanding, not only of those specific systems, but of the structure of complex adaptive systems in general.

Acknowledgements: We thank Rick Riolo for very helpful conversations. This work was supported in part by the US National Science Foundation under grant no. DMI-9908706.

---

[1] D. Challet and Y.-C. Zhang, *Physica A*, **246**, 407 (1997).

[2] R. Savit, R. Manuca and R. Riolo, Phys. Rev. Lett. **82**, 2203 (1999).

[3] R. Manuca, Y. Li, R. Riolo and R. Savit, *The Structure of Adaptive Competition in Minority Games.*, UM Program for Study of Complex Systems Technical Report PSCS-98-11-001, at http://www.pscs.umich.edu/RESEARCH/pscs-tr.html, or LANL eprint at http://xxx.lanl.gov/abs/adap-org/9811005, submitted for publication.

Despite this complication, A(r)'s that favor large minority groups still seem to give the same value of $\sigma^2/N$ as the other (non-normal) games.

[19] The relevant quantity here is, of course, the behavior of $\int rA(r)d\mu(r)$.

## Figure Captions

Fig. 1. $\sigma^2/N$ as a function of $z=2^m/N$ for standard minority game. Each point represents an average value over 32 runs of 10,000 time steps apiece. Triangles indicate the results of games played with N=51 agents, squares correspond to N=101 agents, and circles indicate the results of games played with N=201 agents.

Figs. 2a-e. $\sigma^2/N$ as a function of $z=2^m/N$ for several different games with different pay-off functions. Each point represents an average over 8 runs of 10,000 time steps apiece. In all these figures, triangles indicate the results of games played with N=51 agents, squares correspond to N=101 agents, and circles indicate the results of games played with N=201 agents. 2a) $A(r) = e^{-4r}$, 2b) $A(r) = r^{-2}$, 2c) $A(r)=r^{-1}$ (fixed sum game), 2d) $A(r)=r^2$, 2e) $A(r)=e^{0.1r}$.

Figs. 3a-f. Mean wealth of the agents, $\overline{w}$, as a function of z for the games studied in Figs. 2 (Figs. 3a-e), and for the standard minority game (Fig. 3f). Each point represents an average over 8 runs of 10,000 time steps apiece. In all these figures, triangles indicate the results of games played with N=51 agents, squares correspond to N=101 agents, and circles indicate the results of games played with N=201 agents. The solid lines in these graphs are simple analytic estimates of the mean wealth, as described in Section IV. 3a) $A(r) = e^{-4r}$, 3b) $A(r) = r^{-2}$, 3c) $A(r)=r^{-1}$ (fixed sum game), 3d) $A(r)=r^2$, 3e) $A(r)=e^{0.1r}$, 3f) $A(r)$ is that of the standard minority game.

Figs. 4a-e. Scaled standard deviation of agent wealth, $\tilde{\eta} \equiv \frac{\eta}{\overline{w}}\sqrt{N}$, as a function of z for the games studied in Figs. 2 (Figs. 4a-e), and for the standard minority game (Fig. 4f). Because in these figures we plot results for large values of m (up to m=16), each game was run for 50,000 time steps. Each point represents an average over 8 such runs. In all these figures, triangles indicate the results of games played with N=51 agents, squares correspond to N=101 agents, and circles indicate the results of games played with N=201 agents. 4a) $A(r) = e^{-4r}$, 4b) $A(r) = r^{-2}$, 4c) $A(r)=r^{-1}$ (fixed sum game), 4d) $A(r)=r^2$, 4e) $A(r)=e^{0.1r}$, 4f) $A(r)$ is that of the standard minority game.

Figs. 5a-e. Scaled standard deviation of agent wealth, $\eta/\overline{w}$, as a function of z for the games studied in Figs. 2 (Figs. 5a-e), and for the standard minority game (Fig. 5f). In all these figures, triangles indicate the results of games played with N=51 agents, squares

correspond to N=101 agents, and circles indicate the results of games played with N=201 agents. 5a) $A(r) = e^{-4r}$, 5b) $A(r) = r^{-2}$, 5c) $A(r)=r^{-1}$ (fixed sum game), 5d) $A(r)=r^2$, 5e) $A(r)=e^{0.1r}$, 5f) $A(r)$ is that of the standard minority game.

Fig. 6. $\sigma^2/N$ as a function of z for the payoff function $A(r) = 10$ for $r = r_o \equiv n_o/N$, and $A(r) = 1$, otherwise with different choices of $r_o$ corresponding to favored minority groups of various sizes. In all these experiments, N=101. Each point represents an average value over 32 runs of 10,000 time steps apiece. Triangles indicate $n_o$=49, squares indicate $n_o = 45$ and circles indicate $n_o = 43$.

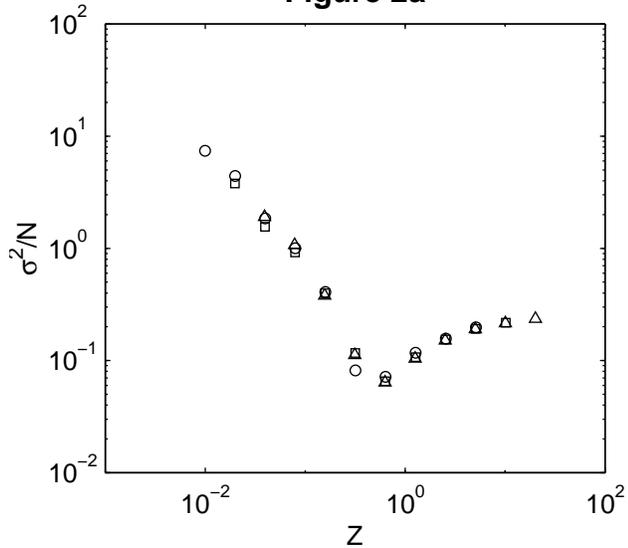

Figure 2a

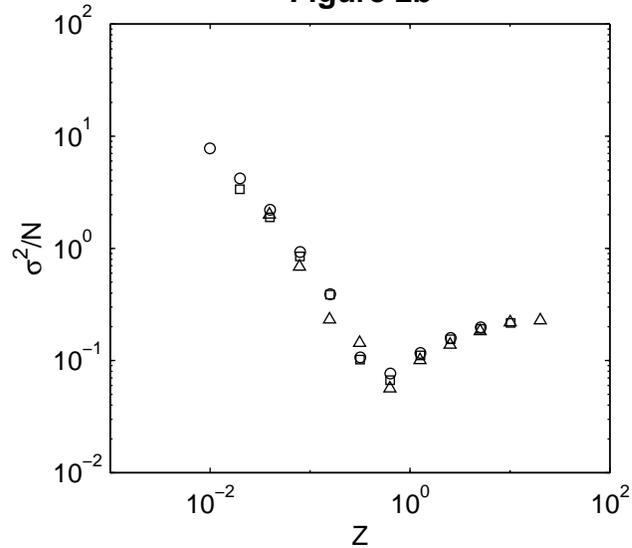

Figure 2b

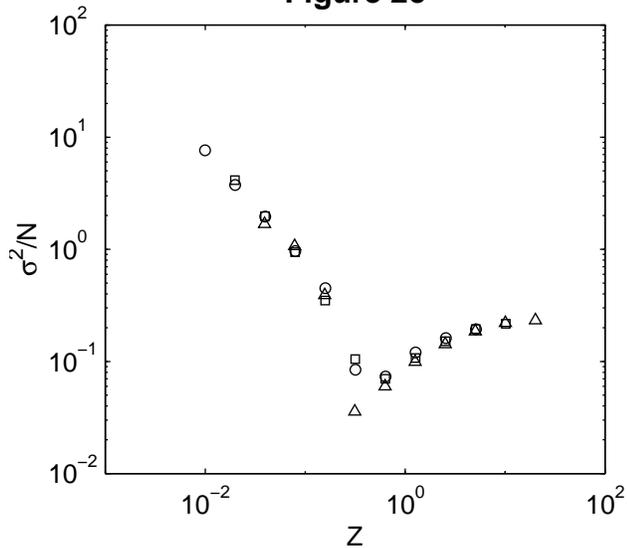

Figure 2c

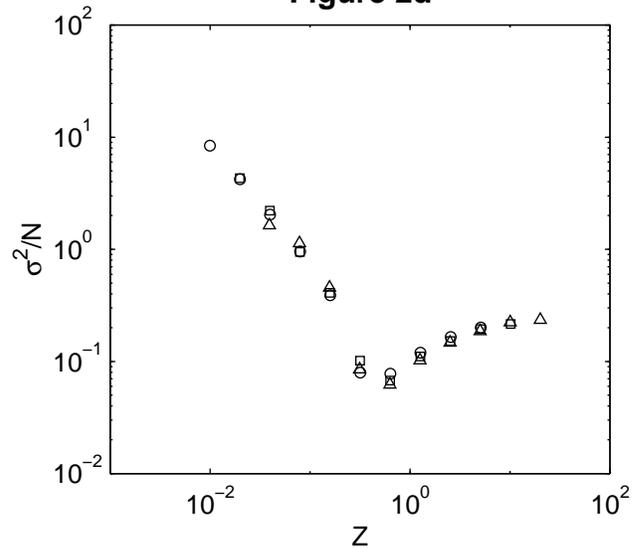

Figure 2d

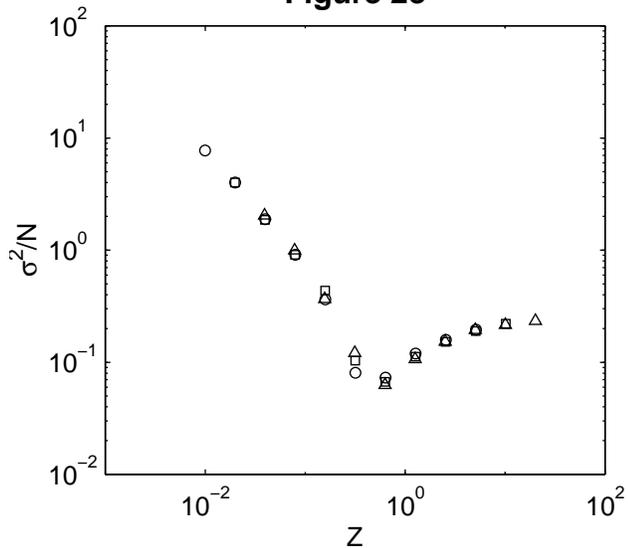

Figure 2e

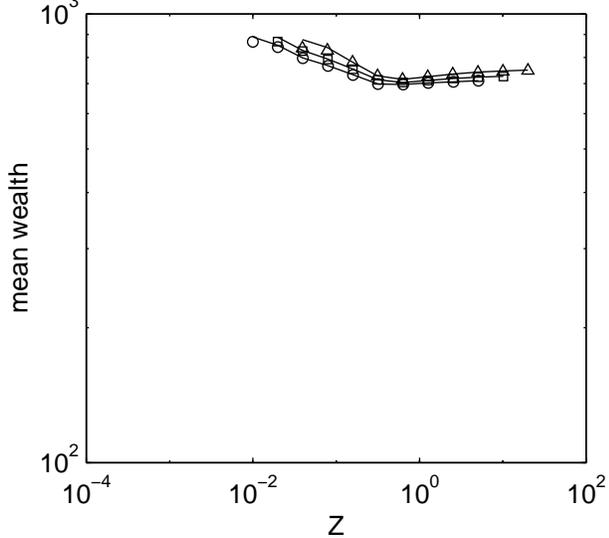
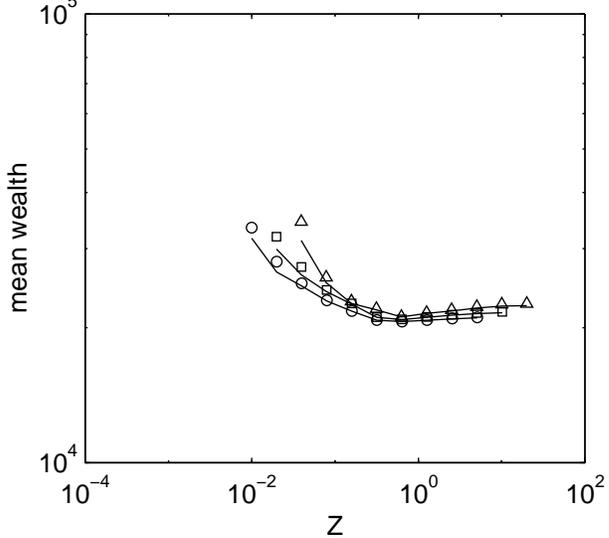
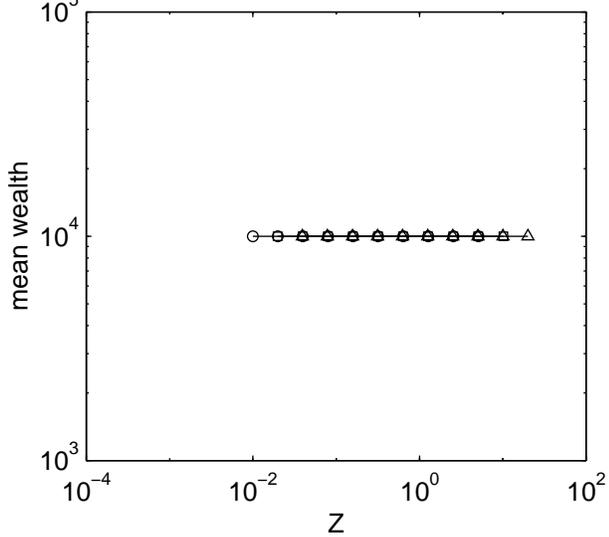
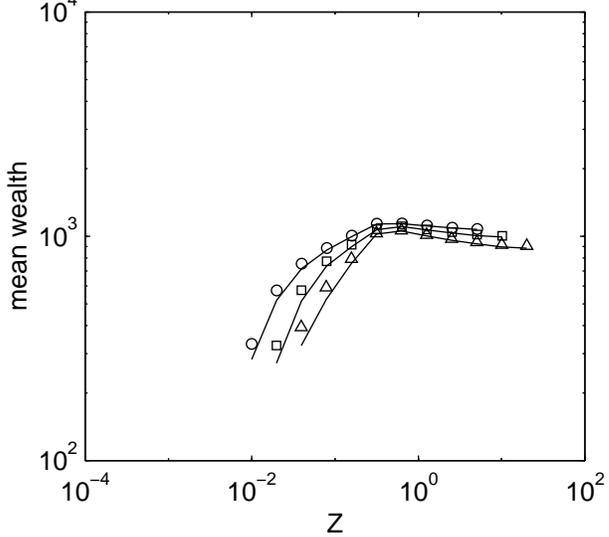
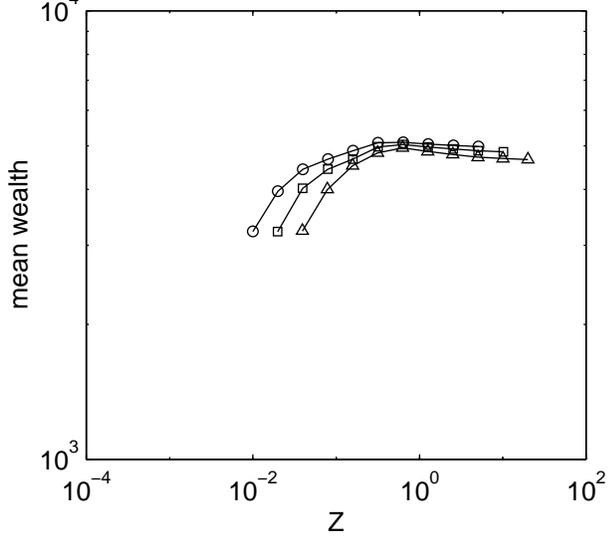
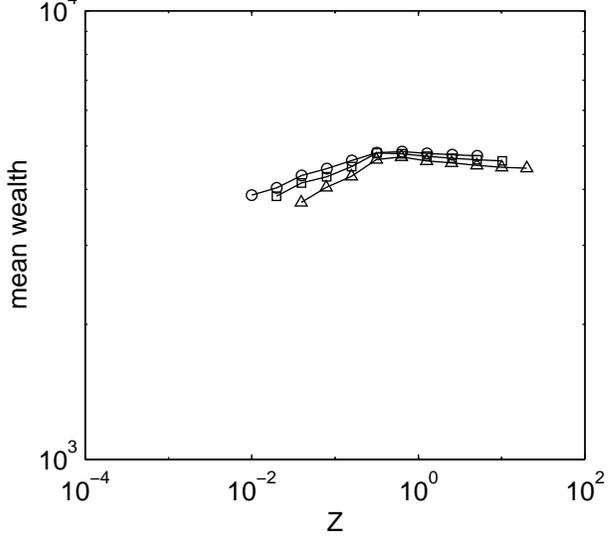

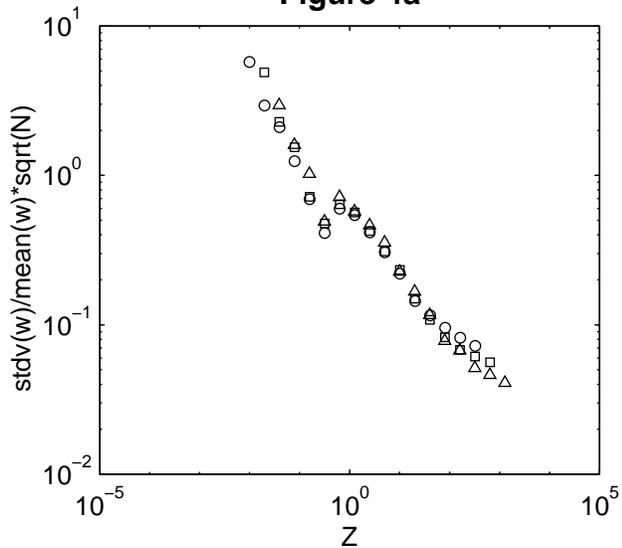
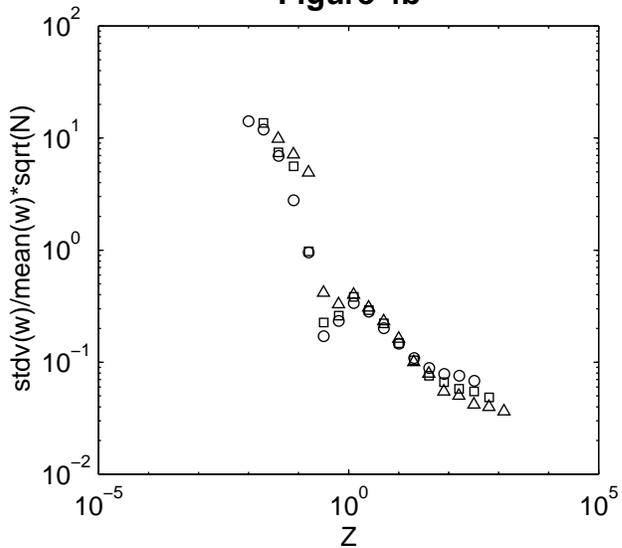
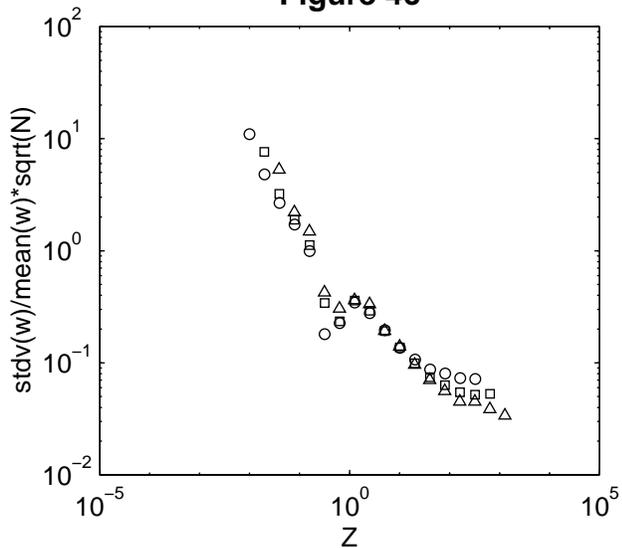
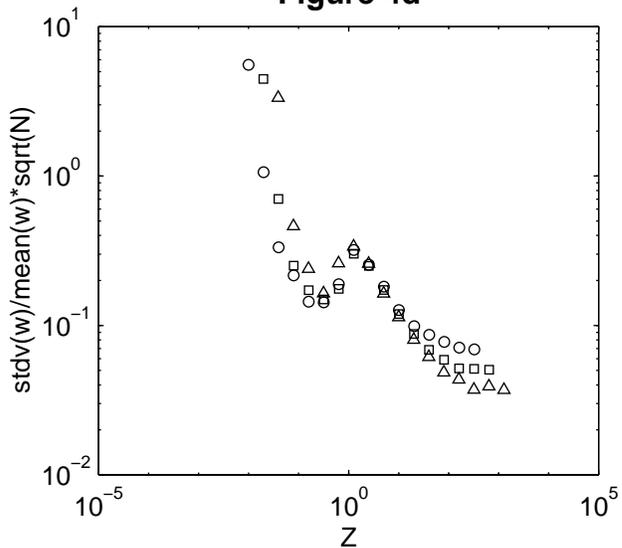
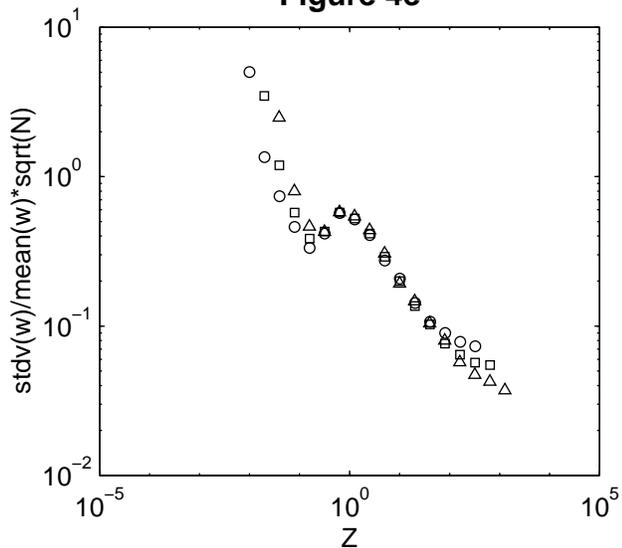
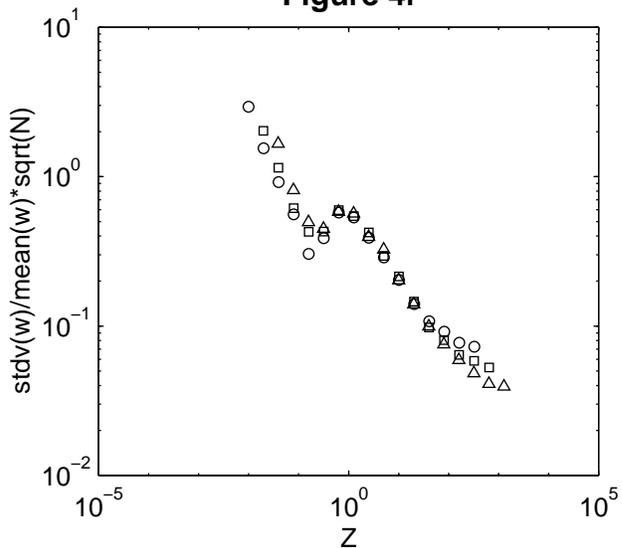

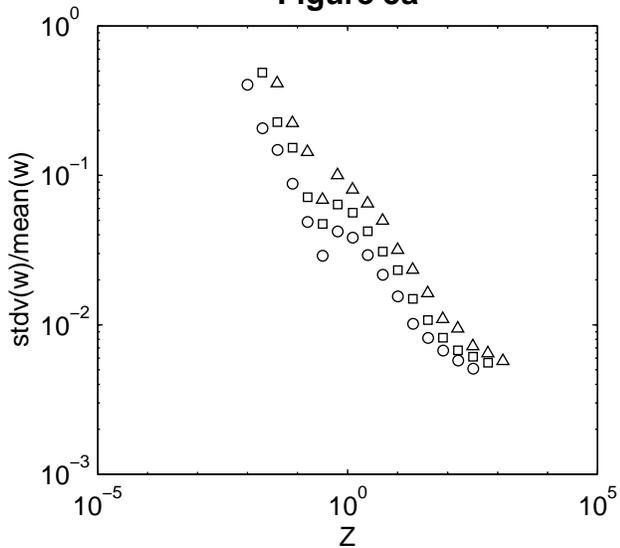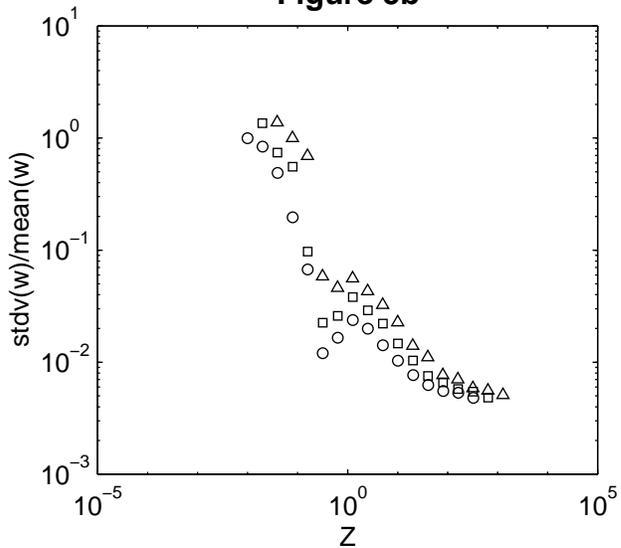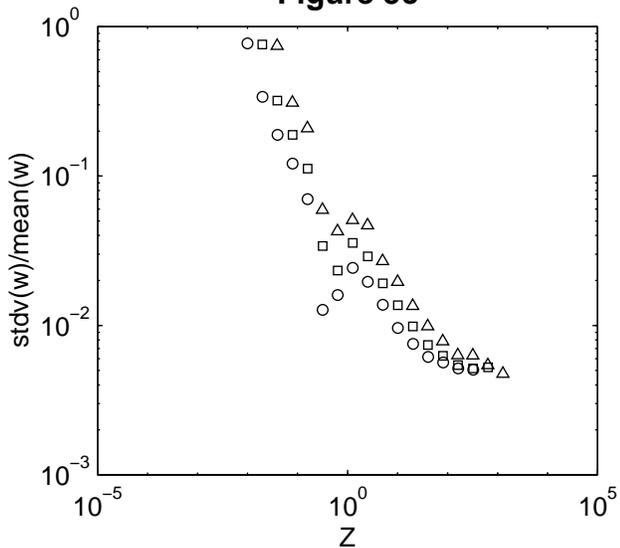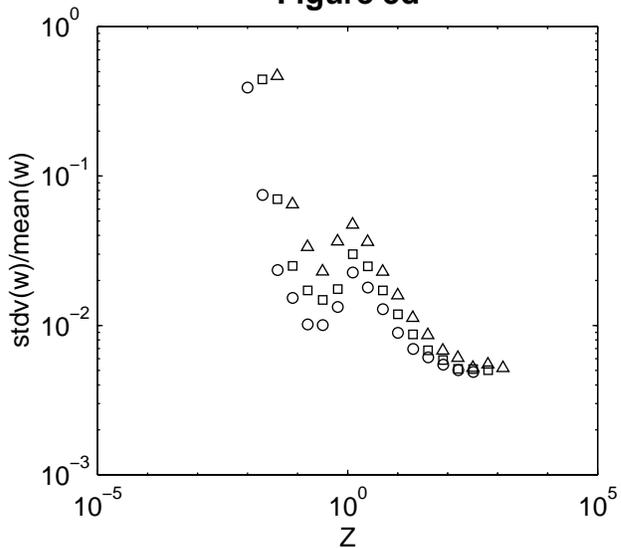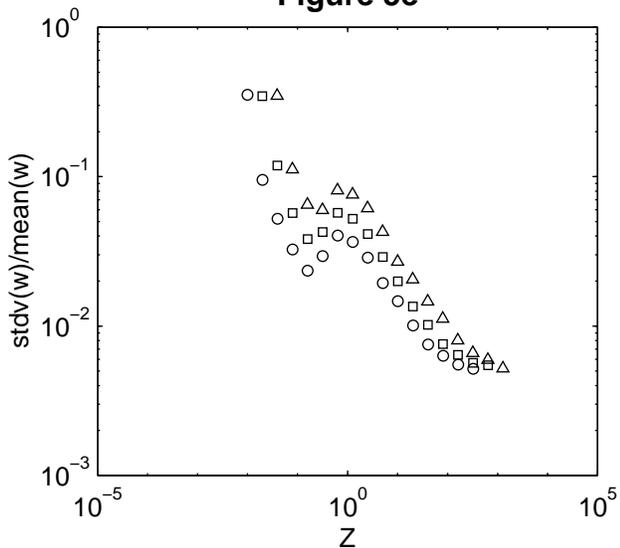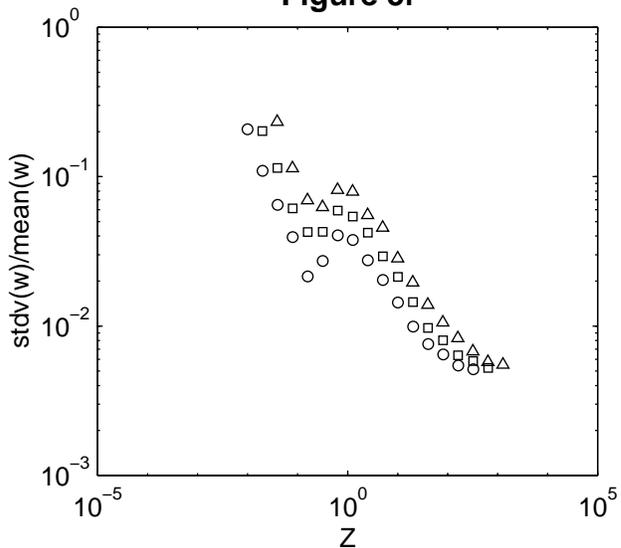

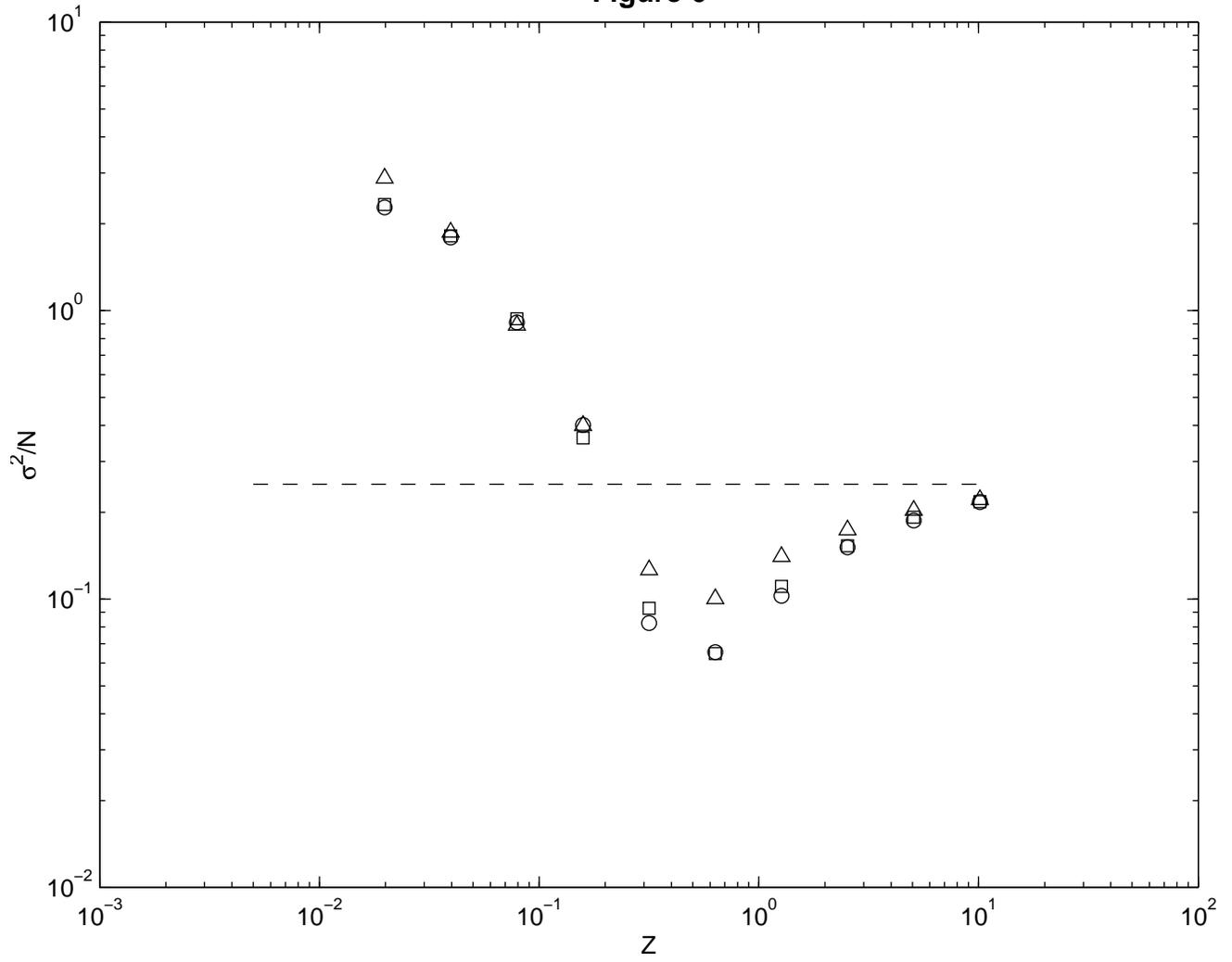

**Figure 6**